\documentclass[preprint,amsmath,showpacs,nofootinbib]{revtex4-1}
\usepackage[T1]{fontenc} 
\usepackage{multirow}
\usepackage{makecell}
\usepackage{longtable}

\newcommand{\sigsip}{\ensuremath{\sigma^{\rm SI}_{\chi p}}}
\newcommand{\sigsdp}{\ensuremath{\sigma^{\rm SD}_{\chi p}}}

\newcommand{\sigxn}{\ensuremath{\sigma_{\chi \mathcal{N}}}}

\newcommand{\sv}{\ensuremath{\langle\sigma v\rangle}}
\newcommand{\dnde}{\ensuremath{\frac{dN_e}{dE_e}}}

\newcommand{\Nfour}{\ensuremath{\frac{dN^{[4 f]}_{\gamma}}{dE_{\gamma}}}}
\newcommand{\Ntwo}{\ensuremath{\frac{dN^{[2 f]}_{\gamma}}{dE_{\gamma}}}}

\newcommand{\mev}{\ensuremath{\,\mathrm{MeV}}}
\newcommand{\gev}{\ensuremath{\,\mathrm{GeV}}}
\newcommand{\tev}{\ensuremath{\,\mathrm{TeV}}}

\newcommand{\rdeg}{\ensuremath{^{\circ}}}

\usepackage{amsmath}
\usepackage{graphicx,bm}
\usepackage[colorlinks,citecolor=blue]{hyperref}
\usepackage[caption=false]{subfig}
\usepackage{slashed}
\usepackage{ulem}

\usepackage{soul}

\begin{document}
\title{Multi-frequency test of dark matter annihilation into long-lived particles in Sirius}

\author{Yu-Xuan Chen$^{a,b}$}
\author{Lei Zu$^{a}$}
\author{Zi-Qing Xia$^{a}$}
\author{\\Yue-Lin Sming Tsai$^{a,b}$}
\author{Yi-Zhong Fan$^{a,b}$}

\affiliation{$^{a}$Key Laboratory of Dark Matter and Space Astronomy, 
Purple Mountain Observatory, Chinese Academy of Sciences, Nanjing 210033, China}
\affiliation{$^{b}$School of Astronomy and Space Science, University of Science and Technology of China, Hefei, Anhui 230026, China}

\begin{abstract}
New long-lived particles produced at the colliders may escape from conventional particle detectors. 
Using satellites or ground telescopes, we can detect the photons
generated from the annihilation of the star-captured dark matter into a pair of long-lived particles. 
When the propagation length of these long-lived particles surpasses the interplanetary distance 
between the Sun and Jupiter, it becomes unfeasible to detect such dark matter signals originating from the Sun or Jupiter on Earth. 
Our analysis of the dark matter-induced photons produced by  prompt radiation, inverse Compton scattering, and synchrotron radiation mechanisms 
reveals that a decay length of about $10^{-3}$ pc for long-lived particles is required for maximum detectability. 
We investigate the parameters that allow the long-lived particle's lifetime to be consistent with Big Bang nucleosynthesis 
while also allowing it to escape the confines of our solar system. 
The Sirius system is proposed as a promising target for the indirect detection of such long-lived particles. 
Utilizing the prompt, inverse Compton scattering, and synchrotron radiation, 
upper limits on the dark matter-proton spin-independent and spin-dependent cross section are estimated with 
the Fermi-LAT null-signal observation and the capabilities of the upcoming Square Kilometre Array radio telescope.

\end{abstract}

\date{\today}

\maketitle

\tableofcontents

\newpage

\section{Introduction} 

The existence of dark matter (DM) has been confirmed by observing its gravitational effects, 
but its interaction with the standard model (SM) particles is still unknown. 
Despite various attempts by scientists to detect non-gravitational interactions, no conclusive evidence has been found (see \cite{Fan:2022dck, Zhu:2022tpr} for tentative signals).
This absence of detection has led to some DM candidates, 
in particular the weakly interacting massive particles (WIMPs), being somewhat challenged. 
However, within the WIMP paradigm, the possibility of DM dominantly annihilating into a pair of long-lived particles 
remains a favored option for producing the correct relic density. 
This makes the search for long-lived particles just as important as searching for DM. 
If DM is heavier than long-lived particles, searching for the latter may prove to be easier than searching for DM directly. 
Several beam dump experiments, e.g., FASER~\cite{Feng:2017uoz} and Belle II~\cite{Belle-II:2018jsg}, 
aim to explore the decay length of long-lived particles in the range of $\mathcal{O}(10)$ m to $\mathcal{O}(10^2)$ m.

The stringent constraints from the collider search~\cite{Jaegle:2015fme,BaBar:2014zli,Merkel:2014avp,LHCb:2019vmc,CMS:2016kce,CMS:2018ffd,KLOE-2:2016ydq,KLOE-2:2011hhj,KLOE-2:2012lii,NA482:2015wmo,Bross:1989mp,Riordan:1987aw,Batell:2014mga,Bjorken:1988as,Marsicano:2018krp,Blumlein:2011mv,Blumlein:2013cua,Gninenko:2012eq,ATLAS:2016olj,ATLAS:2016jza} and the DM direct detection (DD)~\cite{PICO:2017tgi,PICO:2015pux,PICO:2016kso,PICO:2019vsc,LUX:2016ggv,LZ:2022ufs,PandaX-II:2016vec,PandaX-II:2016wea,XENON:2017vdw,XENON:2019rxp,XENON:2019zpr,CDEX:2019hzn,CRESST:2015txj,SuperCDMS:2015eex,Behnke:2016lsk,SIMPLE:2014pun,IceCube:2016dgk,Super-Kamiokande:2015xms}  
suggest a small coupling between a long-lived particle
$\phi$ and SM particles. 
When the decay length of $\phi$ is longer than the size of Earth-based detectors, the absence of detection is expected. 
In particular, if $\phi$ only decays into an electron-positron pair, a longer lifetime 
approximately $10^5$ seconds would not affect the successful history of Big Bang Nucleosynthesis (BBN). 
Thus, finding an astrophysical object to test such a long-lived $\phi$ is an interesting topic. 
If a star passes through a DM halo, collisions between dark matter and nuclei within the star can lead to DM particles becoming trapped by the gravitational potential of the star. 
This accumulation of DM can successively annihilate to $\phi$. 
If the decay length of $\phi$ is long enough to escape from the star, the initial decay into an electron-positron pair can be detected by telescopes. 
In our solar system, the Sun and Jupiter are the best targets for decay lengths shorter than 
$\mathcal{O}(5)$~AU~\cite{Busoni:2013kaa,Leane:2017vag,HAWC:2018szf,1912.09373, 2006.04114, 2103.16794,Leane:2021tjj,Bose:2021cou}. 
Due to the difference in the core temperature between Jupiter and the Sun, the gamma-ray observation of Jupiter is capable of 
exploring DM mass in the MeV scale~\cite{Leane:2021tjj}, 
whereas the Sun is limited to DM mass above $\gev$~\cite{Leane:2017vag,2103.16794}. 
In addition to gamma-ray telescopes, DM-induced gamma-ray signals from dwarf galaxies can be detected~\cite{Bhattacharjee:2020phk} and the time-dependent DM signals can be examined through cosmic ray (CR) telescopes~\cite{Zakeri:2021cur}.

In order to study the effect of $\phi$ with a decay length greater than the scale of the solar system, it is necessary to observe stars outside of our solar system. 
The Sirius system, which is located 2.7 pc away and is the 7th nearest stellar system, offers an ideal opportunity for such observations. This system is composed of a main sequence star known as Sirius A and a white dwarf known as Sirius B, 
which has a 50.13-year orbit~\cite{2017ApJ...840...70B}. 
Sirius A is the brightest star in the sky and emits strong starlight that may interact with high-energy electrons and positrons ($e^\pm$) created by $\phi$, leading to the production of high-energy gamma-rays via inverse Compton scattering (ICS). 
On the other hand, Sirius B is the brightest and nearest white dwarf and 
is known to have a surface magnetic field strength of $10^6$ Guass~\cite{1947Natur.159..658B, 2000PASP..112..873W}. 
This large magnetic field may interact with $e^\pm$ generated by DM annihilation, 
resulting in a significant amount of synchrotron emission in the radio band.

In this paper, we study DM annihilation to long-lived particles from the Sirius system.  
Our approach involves searching for those DM-induced photon signatures in multi-frequency bands. 
The Sirius system is believed to have captured DM from the Milky Way halo and 
the captured DM inside the star undergoes annihilation to long-lived particles, 
which can escape from the surface of the star and eventually decay into SM particles. 
For the gamma-ray band, we use the observation of the Large Area Telescope aboard the Fermi Space Telescope Mission (called Fermi-LAT), 
which is the high-performance gamma-ray telescope for photon energies between 30 MeV and 1 TeV~\cite{2106.12203}.
In the radio band, we consider the upcoming Square Kilometre Array (SKA)~\cite{Braun:2019gdo} 
for the expected sensitivity of DM annihilation in Sirius. 
The SKA is designed with unprecedented sensitivity at centimeter and meter wavelengths, 
which is able to detect the secondary signal from DM~\cite{Chen:2021rea,Beck:2019ukt,Pinetti:2019ztr,Kar:2019cqo,Cembranos:2019noa}.

The structure of the paper is as follows: 
In Sec.~\ref{sec:DM}, the mechanism of DM captured by the Sirius system and its annihilation into two long-lived particles is outlined. 
The energy spectra of photons and $e^\pm$ generated by the DM annihilation process ($\chi\chi\to\phi\phi\to 4e$) are calculated.
Sec.~\ref{section:sec3} covers the recalculation of photon flux generated by DM annihilation 
through three different processes (prompt production, ICS, and synchrotron). 
After examining the propagation of the long-lived particle and CRs, 
Sec.~\ref{sec:tot_spectra} presents the prediction of photon spectra based on four benchmark points. 
The upper limit from the Fermi-LAT observation and the prospective sensitivity for the SKA of the DM-proton cross-section are discussed in Sec.~\ref{sec:result}.
Finally, the conclusion and summary of findings are presented in Sec.~\ref{sec:summary}.

\section{Dark matter capture and annihilation} 
\label{sec:DM}
As a DM $\chi$ hits the star, the momentum would be exchanged between 
DM particle and the nuclei in the stars.  
If the momentum of $\chi$ is larger than the temperature of a star, 
$\chi$ will lose its momentum and then be gravitationally captured by a star. 
On the other hand, if DM mass $m_\chi$ is smaller than $\mathcal{O}(\gev)$, 
the captured DM inside the core of the stars will be accelerated and ejected (a process known as evaporation). This process is much complicated and beyond the scope of this paper. 
In this work, we only focus on the mass range $4\gev<m_\chi<1\tev$ to neglect the effect of evaporation~\cite{Griest:1986yu,Garani:2021feo,Gould:1987ju,Fan:2011dw,Busoni:2013kaa,Busoni:2017mhe,Garani:2017jcj,Bell:2021fye}.  

\subsection{Dark matter captured by Sirius A and Sirius B}
DM in the galactic halo would be captured and accumulated around astrophysical massive objects, 
such as stars, white dwarfs, and neutron stars. 
DM scatters with the nuclei and loses energy as it passes through the star. 
Once the velocity of DM is less than the escape velocity of the star, the DM particle is captured. 
Multiple scatterings between DM and nucleons within the star may occur~\cite{Bramante:2017xlb,Dasgupta:2019juq}.
The probability for a DM particle scattering with the matter inside Sirius A and Sirius B after $k$-th collision is given by~\cite{Bramante:2017xlb} :
\begin{eqnarray}
p_k(\tau)=2\int_{0}^{1} dy\frac{\left(y\tau\right)^k\times e^{-y\tau}}{k!}  y,
\label{eq:pN}
\end{eqnarray}
where $\tau$ is the optical depth of DM, determined by the ratio between DM-nuclei $\sigxn$ and saturation cross section $\sigma_{\rm sat}$, namely $\tau=1.5\times\sigxn/\sigma_{\rm sat}$.\footnote{The factor $1.5$ 
owes to that the optical depth is normalized to one when one DM particle passes through the star, 
see Eq. (14) of Ref.~\cite{Bramante:2017xlb}.} 
For a star with radius $R$, we have $\sigma_{\rm sat}=\pi R^2/N_n$, where $N_n$ is the total number of nucleons in the star. 
By simply assuming that Sirius A entirely consists of hydrogen, 
the total number of nucleons inside Sirius A is $N_n^{\rm SA}=M_{\rm SA}/m_{\rm H}$ 
where Sirius A mass is $M_{\rm SA}\approx 2.063~M_\odot$ and   
hydrogen mass is $m_{\rm H}=1.673\times10^{-24}$~g. 
In addition, the radius of Sirius A is about $R_{\rm SA}\approx1.711R_\odot$. 
On the other hand, we assume that Sirius B has a degenerate carbon core accounting for $99\%$ of the total mass, 
which is enclosed by a thin atmosphere of hydrogen~\cite{Bell:2021fye}. 
DM particles can scatter with carbon in a degenerate core. 
We take the Sirius B mass as $M_{\rm SB}\approx 1.018~M_\odot$, and its radius as $R_{\rm SB}\approx0.0084~R_\odot$. 
Hence, the total number of nucleons inside Sirius B is $N_n^{\rm SB}=0.99~M_{\rm SB}/m_{\rm C}$, 
where carbon mass is $m_{\rm C}=1.993\times10^{-23}$~g.

The cross-section $\sigxn$ can be simply written as the DM-proton spin-independent component $\sigsip$ and 
the spin-dependent component $\sigsdp$,   
\begin{eqnarray}
\sigxn^{\rm SI}&\equiv& \sigsip\times \mathcal{A}^2 \left(\frac{m_\mathcal{N}}{m_p}\right)^2 
\left(\frac{m_\chi+m_p}{m_\chi+m_\mathcal{N}} \right)^2, \\
\sigxn^{\rm SD}&\equiv& \sigsdp \times \frac{4(J_\mathcal{N}+1)}{3 J_\mathcal{N}}\times 
\left| \langle S_p \rangle + \langle S_n \rangle\right|^2 \times
\left(\frac{m_\mathcal{N}}{m_p}\right)^2 
\left(\frac{m_\chi+m_p}{m_\chi+m_\mathcal{N}} \right)^2, 
\label{eq:si_sd}
\end{eqnarray}
where $\mathcal{A}$ is the mass number, $m_p$ and $m_\mathcal{N}$ are proton and nucleon mass, 
$J_\mathcal{N}$ is the nuclear spin, and 
$\langle S_p \rangle$ and $\langle S_n \rangle$ are the expectation values of the
spins of the proton and neutron subsystems~\cite{Bednyakov:2004xq}. 
For the Hydrogen case, one can find $\sigxn\sim\sigsip\sim\sigsdp$. 
However, there is no spin-dependent component for the Carbon case, 
because the nuclear spin of the carbon nucleus is zero, namely $\sigxn\sim 144\sigsip$. 

We assume DM velocities follow the Maxwell Boltzmann distribution but neglect the small systematic uncertainty of the capture rate 
arising from various DM velocity distributions~\cite{Bose:2022ola}.  
By integrating over the DM velocity distribution~\cite{Bramante:2017xlb}, 
the DM capture rate after $k$-th scattering can be written as
\begin{eqnarray}
C_k=\pi R^2 \times p_k(\tau) \times \frac{\sqrt{6}n_\chi}{3\sqrt{\pi}\bar{v}} \times
\left[(2\bar{v}^2+3v_{\rm esc}^2)-(2\bar{v}^2+3v_k^2)\exp{\left(-\frac{3(v_k^2-v_{\rm esc}^2)}{2\bar{v}^2}\right)}\right], 
\label{eq:CRN}
\end{eqnarray}
where we take the DM velocity dispersion as $\bar{v}\sim 270~{\rm km/s}$.  
The escape velocity of Sirius A and Sirius B are $v_{\rm esc}^{\rm SA}=\sqrt{2G_{\rm N}M_{\rm SA}/R_{\rm SA}}\sim 678~{\rm km/s}$  
and $v_{\rm esc}^{\rm SB}=\sqrt{2G_{\rm N}M_{\rm SB}/R_{\rm SB}}\sim 6800~{\rm km/s}$, respectively. 
Here, $G_{\rm N}$ is the gravitational constant.  
The velocity of DM particle undergoing $k$ scatters is $v_k=v_{\rm esc}(1-\beta_{+}/2)^{-k/2}$ 
where we define $\beta_{+}\equiv 4m_\chi m_\mathcal{N}/(m_\chi + m_\mathcal{N})^2$~\cite{Bramante:2017xlb} 
with the DM and nucleus mass defined as $m_\chi$ and $m_\mathcal{N}$. 
We can obtain the DM number density near Sirius A and Sirius B by $n_\chi=\rho_\chi/m_\chi$. 
Since the Sirius system is located around $2.7$~pc from the Earth, 
we take the local density $\rho_{\chi}\sim 0.4\gev/\rm{cm}^3$ as an approximation. 
Finally, the total capture rate of DM in Sirius A or Sirius B is the sum of all individual $C_k$,
\begin{eqnarray}
C_{\rm tot}=\sum_{k=1}^\infty C_k.
\label{eq:CRtot}
\end{eqnarray}
We cut off the sum in Eq.~\eqref{eq:CRtot} when $p_k(\tau)$ does not change by increasing $k$.

\begin{figure*}[!ht]
\centering 
\includegraphics[width=0.49\textwidth]{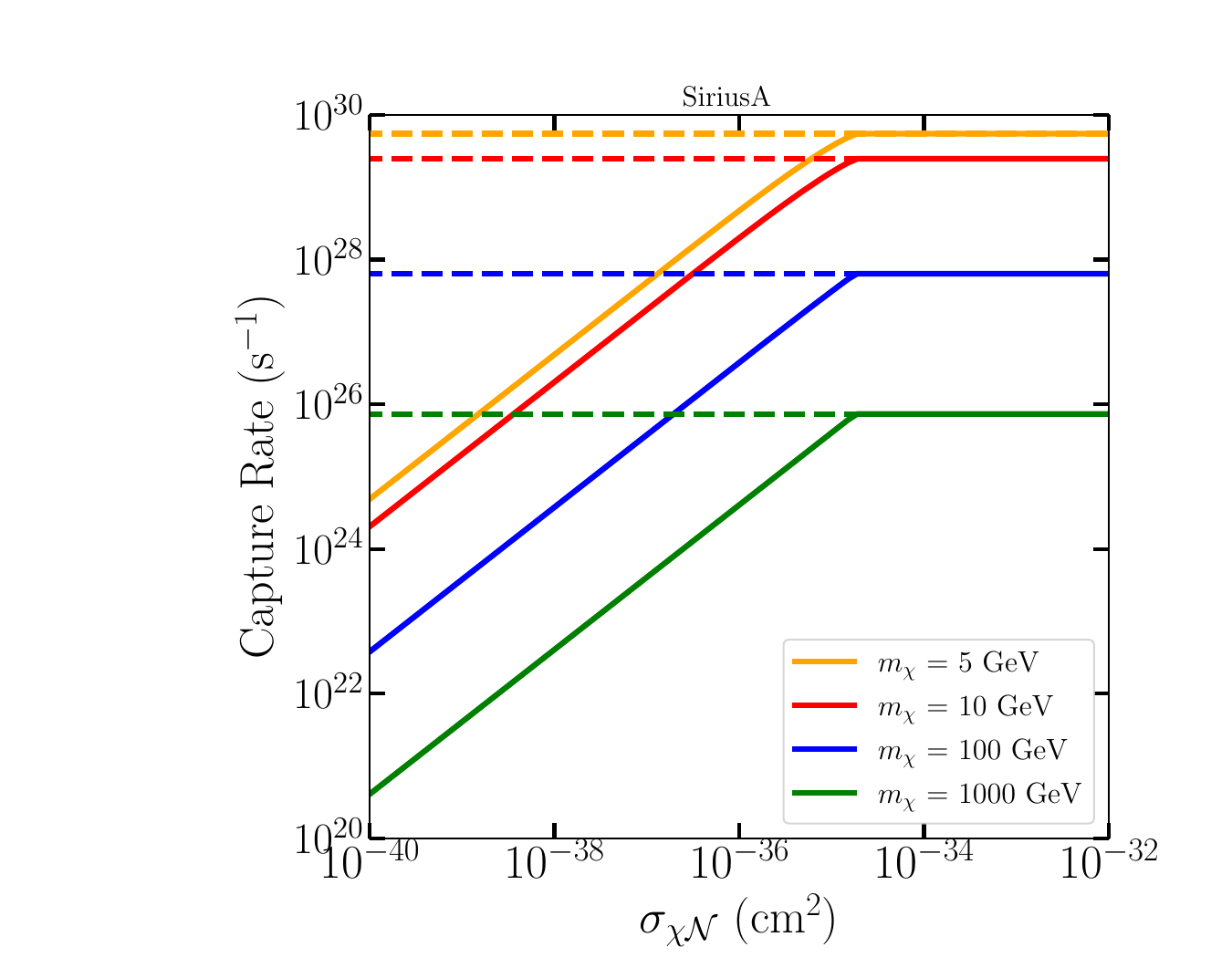}
\includegraphics[width=0.49\textwidth]{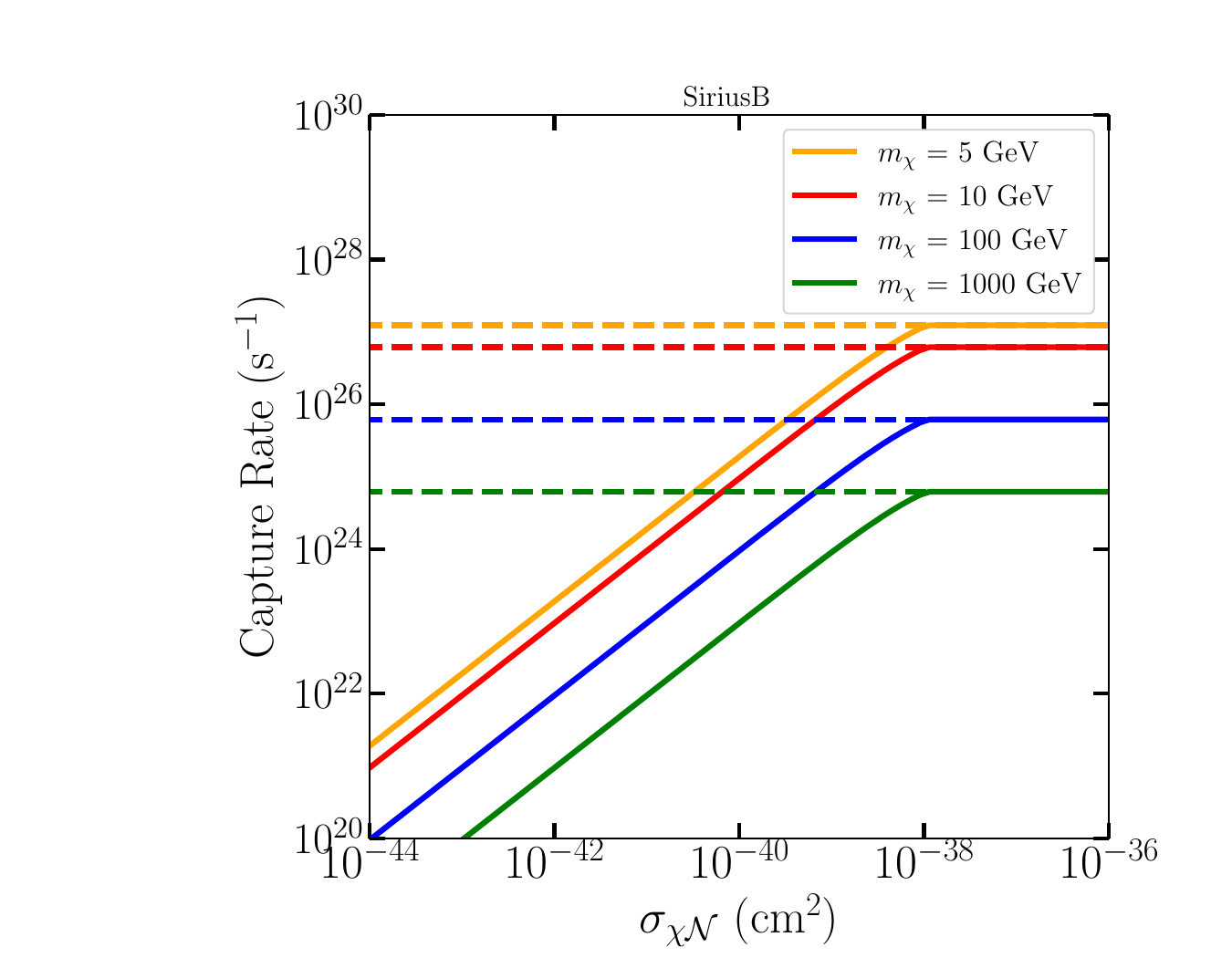}\\
\caption{Capture rates as a function of the DM-nuclei scattering cross section $\sigxn$ for 
four different DM masses $m_{\chi} = 5,~10,~100,~1000$ GeV. 
The left and right panels correspond to the case of Sirius A and B, respectively. 
The dashed lines indicate the absolute geometric upper limits, while  
the solid lines include the effects of multiple scattering. 
Note that the capture rate is truncated once it exceeds the corresponding geometric limit.} 
\label{Fig:Capture_Rate} 
\end{figure*}

Fig.~\ref{Fig:Capture_Rate} shows DM model-independent capture rates of Sirius A and Sirius B
as a function of $\sigxn$ for Sirius A (left panel) and B (right panel). 
The different values of $m_\chi$ are presented as different colors.   
Note that we have demonstrated the geometric upper limits (maximal capture rates) by using the dashed lines.    
We find that the DM capture rate decreases with increasing $m_\chi$. 
This is because the number density of DM $n_\chi$ is inversely proportional to $m_\chi$ for a given local mass density. 
We also see that the differences of the maximal capture rates for Sirius A and B are roughly 
from $\mathcal{O}(10)$ to $\mathcal{O}(10^3)$, regardless of $m_\chi$. 
Thus the expected maximal gamma-ray flux from Sirius B is far less than that from Sirius A for ICS and prompt production. 
We will discuss the DM-induced photon flux in Sec.~\ref{section:sec3}.

\subsection{Dark matter annihilation rate}
When DM particles are captured in Sirius A and Sirius B, they will annihilate to pairs of long-lived particles. 
Considering the number of DM particles $N(t)$ inside Sirius A or Sirius B at the time $t$, 
we have~\cite{Press:1985ug,Griest:1986yu,Silk:1985ax,Krauss:1985ks,Gould:1987ju,Gould:1987ir,Jungman:1995df,Peter:2009mk,Zentner:2009is,Busoni:2013kaa,Busoni:2017mhe,Garani:2017jcj,Nguyen:2022zwb,Bose:2021yhz,Bhattacharjee:2022lts}
\begin{eqnarray}
\Dot{N}(t)=C_{\rm cap}-2\Gamma_{\rm ann}(t) -C_{E} N(t),
\label{eq:Ndott}
\end{eqnarray}
where $C_{\rm cap}$ is the total DM capture rate given by Eq.~\eqref{eq:CRtot} and 
demonstrated in Fig.~\ref{Fig:Capture_Rate}.
When the DM momentum is below the star temperature, DM particles would be evaporated.  

On the other hand, in the DM mass range of our interest $m_\chi>4\gev$,  
we can safely take $C_{E}=0$ for both Sirius A and Sirius B. 
This is because that Sirius A is similar to the Sun\footnote{Actually, the DM evaporation mass of Sirius A is slightly larger than the sun~\cite{Garani:2021feo} due to its slightly higher temperature, which has negligible impact for our work.}, which DM evaporation mass is $4~\gev$~\cite{Griest:1986yu,Gould:1987ju,Busoni:2013kaa,Busoni:2017mhe,Garani:2017jcj}. 
As for Sirius B (the white dwarf), evaporation becomes significant only if $m_\chi<\mathcal{O}(\mev)$~\cite{Garani:2021feo,Bell:2021fye}.

If DM annihilates after being thermalized, DM particles will follow the Boltzmann distribution in the core of the star with the number density
\begin{eqnarray}
n_\chi(r)=n_0 e^{-r^2/r_\chi^2},~~{\rm where} 
\qquad
r_\chi=\sqrt{\frac{3T_{\rm c}}{2\pi G_{\rm N} \rho_{\rm c} m_\chi}}\approx0.01R \sqrt{\frac{100 \rm GeV}{m_\chi}}.
\label{eq:nchir}
\end{eqnarray}
In this equation, $T_{\rm c}$ and $\rho_{\rm c}$ are the temperature and density of the star center, respectively. 
The radius of the star is represented by $R$, and the gravitational constant is represented by $G_{\rm N}$. 
The annihilate rate is then written as 
\begin{eqnarray}
\Gamma_{\rm ann}(t)=\frac{1}{2}C_{\rm ann}N(t)^2, 
\label{eq:Gamma}
\end{eqnarray}
where $C_{\rm ann}$ is the DM self-annihilation rate~\cite{Feng:2016ijc,Baratella:2013fya} 
\begin{eqnarray}
C_{\rm ann}=\sv \left(\frac{G_{\rm N} m_\chi \rho_{\rm c}}{3T_{\rm c}}\right)^{3/2}, 
\label{eq:Cann}
\end{eqnarray}
where $\sv$ is the velocity averaged annihilation cross section. 
The solution of Eq.~\eqref{eq:Ndott} is 
\begin{eqnarray}
N(t)=\sqrt{\frac{C_{\rm cap}}{C_{\rm ann}}}\tanh{\frac{t}{\tau_{\rm eq}}},
\label{eq:Nt}
\end{eqnarray}
and $\tau_{\rm eq}=1/\sqrt{C_{\rm cap} C_{\rm ann}}$ is the time scale
to characterize the equilibrium $\Dot{N}(t)=0$. 
We take $T_{\rm c}^{\rm SA}=3.5\times 10^7 ~\rm K$ and $\rho_{\rm c}^{\rm SA}\sim \mathcal{O}(100)~\rm {g/cm^3}$ for Sirius A, 
as well as $T_{\rm c}^{\rm SB}\sim \mathcal{O}(10^9) ~\rm K$ and $\rho_{\rm c}^{\rm SB}\sim \mathcal{O}(10^6)~\rm {g/cm^3}$ 
for Sirius B. 
Therefore, we obtain the equilibrium time scales $\tau_{\rm eq}^{\rm SA}\approx 2\times10^{8}$~yrs and 
$\tau_{\rm eq}^{\rm SB}\approx 2.5\times10^{6}$~yrs based on the DM parameters: $(\sv\sim 10^{-26}$~cm$^3$~s$^{-1}$, $C_{\rm cap}\sim 10^{23}$~s$^{-1}$, $m_\chi \sim 100 \gev)$. 
These are shorter than the life of the Sirius system $\mathcal{O}(3\times10^8)$~yrs. We emphasize that the equilibrium assumption holds for most parameter spaces of our interest. Therefore, the annihilation rate $\Gamma_{\rm ann}$ after reaching the equilibrium state is then simply
\begin{eqnarray}
\Gamma_{\rm ann}=\frac{1}{2}C_{\rm cap},
\label{eq:Gamma}
\end{eqnarray}
where the factor $1/2$ accounts for two DM particles participating in the annihilation.

\subsection{Dark matter annihilation to a pair of long-lived particles}
Considering DM particles $\chi$ annihilates into a pair of long-lived particles $\phi$ and then decays to a pair of SM particles $f\bar{f}$, where $f$ represents the fermion in SM-like $e^\pm$, $\mu^\pm$, $b\bar{b}$ and so on.  
Therefore, DM annihilation leads to four SM particles $4f$ in the final state. 
The secondary photons generated from this process are able to be detected by multi-frequency observation. 
In the following example, we take the photon energy spectrum in the lab frame, 
\begin{eqnarray}
\Nfour(m_\chi, m_\phi, E_{\gamma})=
\int_{E_{\rm min}}^{E_{\rm max}} dE_f \frac{dN_{f}}{dE_{f}}(m_\chi, m_\phi, E_{f}) 
\Ntwo(E_f, E_{\gamma}). 
\label{eq:dnde}
\end{eqnarray}
The box-shape energy spectrum $dN_f/dE_f$ describes the process of DM annihilation  $\chi\chi\to\phi\phi\to 4 f$ , 
\begin{eqnarray}
\frac{dN_f}{dE_f}(m_\chi, m_\phi, E_f)= \frac{4}{E_{\rm max}-E_{\rm min}} 
\Theta(E_f-E_{\rm min})\Theta(E_{\rm max}-E_f), 
\label{eq:box}
\end{eqnarray}
where $\Theta$ is the Heaviside function. The maximum and minimum $E_f$ are 
\begin{eqnarray}
 E_{\rm max/min}&=& \frac{m_\chi}{2} \pm 
\left(m_\chi \sqrt{1-\frac{m_\phi^2}{m_\chi^2}}  \sqrt{\frac{1}{4}-\frac{m_f^2}{m_\phi^2}}  \right). 
\label{eq:Emuon}
\end{eqnarray}
We take the spectrum $\Ntwo(E_f, E_{\gamma})$ from \texttt{PPPC4}~\cite{1012.4515,1009.0224} 
by inserting the central energy equal to $2 E_f$ instead of $2 m_\chi$.

\section{The propagation of long-lived particles} 
\label{section:sec3}

\begin{figure}
\includegraphics[width=0.49\textwidth]{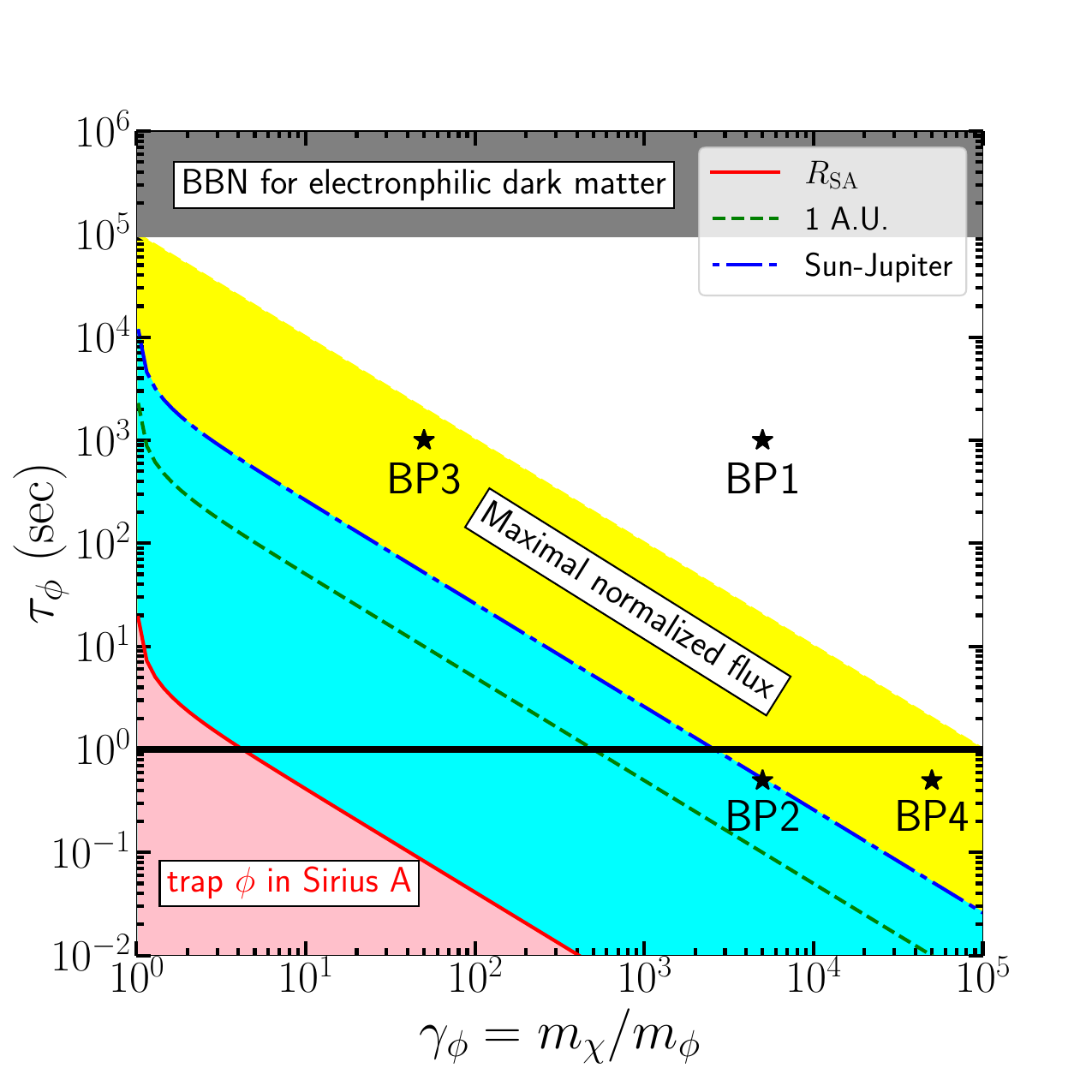}
\includegraphics[width=0.49\textwidth]{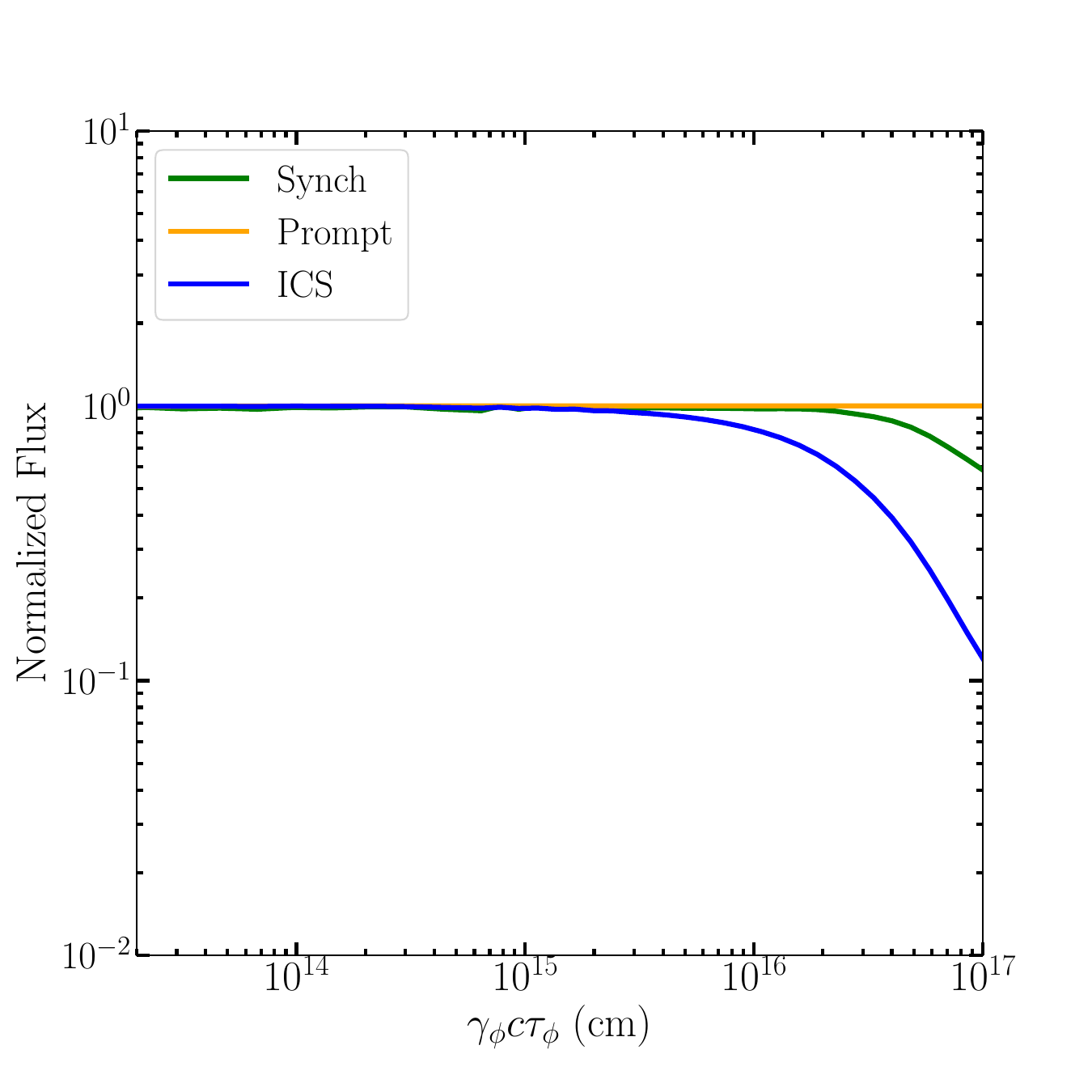}
\caption{Left panel: the parameter space on the ($\gamma_\phi$, $\tau_\phi$) plane 
considered in this work. 
The pink region shows the case that $\phi$ cannot escape from Sirius A, 
while the cyan region is the $\phi$ propagation length shorter than the distance 
between the Sun and Jupiter.  
At the cyan region, \cite{Leane:2021tjj} suggests to detect the long-lived particles  
by using the Jupiter gamma-ray observation. 
The gray region indicates that $\phi$ entirely decays to two electrons after BBN epoch and 
alters the successful BBN history. 
The yellow region shows the corresponding $\gamma_\phi c\tau_\phi$ required by the maximal normalized photon fluxes.  
Right panel: Normalized photon fluxes of Sirius A as a function of $\gamma_\phi c \tau_\phi$ of mediators. 
The blue line is for ICS including the effects of the Cosmic Microwave Background (CMB) and the starlight. 
The green line is for the synchrotron radiation and the orange line is for the final state radiation, respectively. 
In this work, we adopt $\gamma_\phi c \tau_\phi$ between $1.19\times10^{11}~\rm cm$ (the radius of Sirius A) and $3\times10^{15}~\rm cm$ 
for the maximal normalized photon fluxes, see the yellow region in the left panel. 
}
\label{fig:ga_tau}
\end{figure}

Usually, the upper limit of the $\phi$ lifetime $\tau_\phi$ is determined by 
the BBN limit~\cite{1605.07195}. 
If $\phi$ decays to baryon, 
the lifetime $\tau_\phi$ should be shorter than one second depending on 
the BBN measurements (the black solid line in the left panel of Fig.~\ref{fig:ga_tau}). 
However, if $\phi$ only decays to an electron-positron pair, the upper limit of a lifetime from BBN history becomes much weaker, 
namely $\tau_\phi<10^{5}$ seconds as the region below the gray region in the left panel of Fig.~\ref{fig:ga_tau}. 
In this work, we simply respect these BBN limits but focus on 
the case of the decay length of $\phi$ longer than the distance between the Sun and Jupiter (cyan region).
Taking $1<\tau_\phi/{\rm s}<10^{5}$, 
the $\mathcal{O}(10)~\mev$ $\phi$ with a Lorentz boost factor $\gamma_\phi\sim 10^5$ generated by 
DM annihilation in TeV scale, would travel from $\mathcal{O}(1)$~a.u. to $\mathcal{O}(1)$~pc. 

If the decay length $\gamma_\phi\beta\tau_\phi$ exceeds the radius of Sirius A ($R_{\rm SA}=1.7711 R_{\odot}$), 
the long-lived particles would escape from Sirius A with the condition, 
\begin{eqnarray}
\gamma_\phi\beta\tau_\phi=c \tau_\phi\times \sqrt{\frac{m_\chi^2}{m_\phi^2} -1 }  >  1.7711 R_{\odot}.
\label{eq:decay_length}
\end{eqnarray}
In the left panel of Fig.~\ref{fig:ga_tau}, we plot the $\phi$ trapped region (pink) on the ($\gamma_\phi$, $\tau_\phi$) plane. 
We would like to emphasize that the cyan region indicates that the $\phi$ decaying length is shorter than the distance between the Sun and Jupiter. 
Within the cyan region, the most stringent limit is to detect the gamma-ray events from Jupiter 
created by the $\phi$ decay~\cite{Leane:2021tjj}. 
Jupiter is not a good target to detect DM annihilation to a pair of $\phi$ in the yellow region and white region, 
while we can still explore these region through the observation of Sirius.
In this work, we select two benchmark points (BP3 and BP4) from the yellow region, one benchmark point (BP1) from the white region
and also one benchmark point (BP2) from the cyan region as a comparison, which are marked as black stars in the left panel of Fig.~\ref{fig:ga_tau}.

In the right panel of Fig.~\ref{fig:ga_tau}, we plot the normalized fluxes from Sirius A for three distinct components - prompt production (orange line), synchrotron radiation (green line), and ICS (blue line) as a function of the $\phi$ propagation length. 
Here, the normalization factor is the maximum flux by varying the value of $\gamma_\phi c\tau_\phi$.   
We find that the shapes of the normalized fluxes with respect to $\gamma_\phi c\tau_\phi$ 
do not depend on DM parameters such as $m_\chi$ and $\sigxn$.  
In addition, the maximal photon fluxes occur at the region 
$\gamma_\phi c \tau_\phi<3\times 10^{15}~{\rm cm}$ 
if only $\gamma_\phi c\tau_\phi$ is longer than the distance between the Sun and Jupiter (see the yellow region in the left panel of Fig.~\ref{fig:ga_tau}). 
Because the prompt production does not depend on the propagation environment, 
the fluxes are always the same, regardless of $\gamma_\phi c\tau_\phi$. 
On the other hand, the photons produced via the synchrotron or ICS mechanism 
strongly depend on the interaction between $e^\pm$ and the magnetic or photon field.  
Thus, for $\gamma_\phi c\tau_\phi\geq 10^{16}~\rm cm$, 
the $\phi$ decay distance is so large that the small amount of $\phi$ cannot produce enough $e^\pm$ to interact with the magnetic field or photon field, 
resulting in the suppression of photon fluxes eventually.

In the later subsections, we will derive the analytic expression of photon fluxes generated by $\chi\chi\to\phi\phi\to 4 e$ 
via the prompt production (Sec.~\ref{sec:prompt}), ICS (Sec.~\ref{sec:ICS}), and 
synchrotron radiation (Sec.~\ref{sec:syn}).

\subsection{Dark matter prompt gamma-ray emission}
\label{sec:prompt}
Considering that $\phi$ decays outside Sirius A and Sirius B, 
the prompt photon spectrum from the final state radiation arriving at Earth is written as
\begin{eqnarray}
E_{\gamma}^2\frac{d\Phi_\gamma}{dE_\gamma}=\frac{\Gamma_{\rm ann}}{4\pi D_{\rm A}^2} \times E_{\gamma}^2\Nfour \times 
{\rm BR}(\phi\rightarrow f\bar{f}) 
\times \left[\exp\left(-\frac{R}{\gamma_\phi c \tau_\phi}\right)
-\exp\left(-\frac{D_{\rm A}}{\gamma_\phi c \tau_\phi}\right)
\right],
\label{eq:flux_fr}
\end{eqnarray}
where $D_{\rm A}$ is the average distance between the Sirius system and Earth, and 
${\rm BR}(\phi\rightarrow f\bar{f})$ is the branching ratio of the mediator decaying to a pair of $f$ final state. 
In the last two exponential terms, we describe the $\phi$ decaying probability with respect to $\tau_\phi$. 
If decaying length $\gamma_\phi c \tau_\phi$ is much less than $R_{\rm SA}$ ($R_{\rm SB}$), 
we expect a null result since most of $\phi$ decays inside Sirius A (Sirius B). 
Similarly, the condition $ \gamma_\phi c \tau_\phi \le D_{\rm A}$ ensures that $\phi$ would decay before reaching Earth.

\subsection{Diffusion equation}
After DM annihilation to a pair of $\phi$, 
these charge-neutral particles will escape from the star and subsequently decay into relativistic electrons and positrons. 
By scattering with the CMB or starlight, those high-speed electrons and positrons can also produce gamma-ray via the ICS process.  
We take a simple assumption that the electrons and positrons propagate with no convection, 
no reacceleration, and a very large time scale for the fragmentation and radioactive decay.   
Thus we obtain a simple electron and positron propagation equation, 
\begin{equation}
    -\nabla\left[ D(E,\textbf{r}) \nabla \frac{\partial n_e}{\partial E}\right] 
    -\frac{\partial }{\partial E}\left[b(E,\textbf{r})\frac{\partial n_e}{\partial E}\right]
    =Q(E,\textbf{r}), 
    \label{eq:diffep}
\end{equation}
where $\partial n_e/\partial E$ is the equilibrium electron density. We assume equilibrium and hence the equation is time-independent. The energy loss term $b(E, \mathbf{r})$ accounts the total energy loss via synchrotron, IC, Coulomb, and bremsstrahlung processes. The explicit form of the energy loss term is 
\begin{equation}
\begin{aligned}
b(E, \mathbf{r}) &=b_{\rm IC}(E)+b_{\rm syn.}(E, \mathbf{r})+b_{\text {coul. }}(E)+b_{\text {brem. }}(E) \\
&=b_{\rm IC}^{0} E^{2}+b_{\rm syn.}^{0} B^{2}(r) E^{2} +
b_{\text {coul. }}^{0} n_{e}\left[1+\log \left(\frac{E / m_{e}}{n_{e}}\right) / 75\right]\\
& +b_{\text {brem. }}^{0} n_{e}\left[\log \left(\frac{E / m_{e}}{n_{e}}\right)+0.36\right], 
\end{aligned}
\end{equation}
where $n_e$ represents the average thermal electron density. 
In this work, we take $n_e=0.1~{\rm cm}^{-3}$ for the Milky Way~\cite{McDaniel:2017ppt}. 
We follow the Refs.~\cite{Colafrancesco:2006he,Colafrancesco:2014coa,McDaniel:2017ppt}
to take the energy loss factors $b_{\rm IC}^{0}=0.25$, $ b_{\rm syn.}^{0}=0.0254$, 
$b_{\rm coul. }^{0}=6.13$, and $b_{\rm brem. }^{0}=1.51$ in units of $10^{-16}\gev/s$. 
The magnetic field in the propagation path consists of the galactic magnetic field $B_{\rm G}$ and the stellar magnetic field $B_{\rm S}(r)$ which decreases with the distance to the star. One can take $B(r)=B_{\rm S}(r)$ near the star but $B(r)\approx B_{\rm G}$ for $B_{\rm S}(r)\ll B_{\rm G}$ 
as given by~\cite{Wang:2021hfb} 
\begin{eqnarray}
B_{\rm S}(r)\approx B_0\times \left(\frac{R}{r}\right)^3, 
\label{eq:magnetic_field}
\end{eqnarray}
where $B_0$ is the magnetic field strength at the surface of the star in the direction of the magnetic pole~\cite{Cantiello:2019cgv}, 
and $r$ is the distance to the center of the star. We take $B_0^{\rm SA}=0.2$~G for Sirius A and $B_0^{\rm SB}=1$~MG for Sirius B. Noted that we take the axial magnetic field strength as the approximation of the actual magnetic field in Eq.~\eqref{eq:magnetic_field}, 
because $\phi$ propagates a long distance.
The diffusion coefficient $D(E,r)$ can be modeled as 
\begin{equation}
D(E,r)=D_0\frac{d_{\rm B}^{2/3}}{B(r)^{1/3}}\left(\frac{E}{\gev}\right)^{1/3}, 
\end{equation}
where $d_{\rm B}$ is the minimum scale of uniformity of the magnetic field in kpc, 
and we take $d_{\rm B}=20$ for our galaxy. 
According to the measured B/C ratio data in the Milky Way~\cite{Maurin:2001sj,Webber:1992dks}, 
the diffusion coefficient $D_0$ ranges from $10^{27}$ to $10^{29}~{\rm cm}^2 s^{-1}$.

The $\phi$ decay relevant source term $Q(E,r)$ is written as  
\begin{equation}
Q(E,r)=\frac{\Gamma_{\rm ann}}{4{\pi}{r}^2}\times\dnde\times\frac{dP(r)}{dr}, 
\label{source_function}
\end{equation}
where $P(r)=1-\exp{(-r/\gamma_\phi c \tau_\phi)}$ is the $\phi$ decaying probability before reaching $r$, 
so that the average decaying probability of $\phi$ at a small interval $dr$ is
\begin{equation}
\frac{dP(r)}{dr}=\frac{1}{\gamma_\phi c \tau_\phi}\exp\left[-\frac{r}{\gamma_\phi c \tau_\phi}\right].
\end{equation}
The electron spectrum $dN_e/dE_e$ can be obtained by Eq.~\eqref{eq:box} with the replacement $f\to e^\pm$.

We can then solve Eq.~\eqref{eq:diffep} analytically as
\begin{equation}
\begin{split}  
\frac{dn_e}{dE}=\frac{1}{b(E,r)} \int_{E}^{E_{\rm max}} dE' \frac{1}{[4\pi(v-v')]^{1/2}} \sum_{n=-\infty}^{+\infty} (-1)^n \int_{R}^{r_{\rm h}} dr' \frac{r'}{r_n} \\
\cdot \left[\exp{\left(-\frac{(r'-r_n)^2}{4(v-v')}\right)}-\exp{\left(-\frac{(r'+r_n)^2}{4(v-v')}\right)}\right] Q(E',r'), 
\end{split}
\label{ED}
\end{equation}
where $r_{\rm h}$ is the radius of the diffusion zone.
We focus on the diffusion of $e^\pm$ yielded by the decay of $\phi$ between the Earth and Sirius system. Thus, we set $r_{\rm h}=D_{\rm A}$ as the boundary condition by using the image charges method and introducing the charges placed at $r_n=(-1)^nr+2nr_{\rm h}$. 
Here, the parameter $v$ characterizes the diffusion coefficient and energy loss term, 
\begin{equation}
v(E,r)=\int_{E}^{E_{\rm max}} d\Tilde{E} \frac{D(\Tilde{E},r)}{b(\Tilde{E},r)}.
\label{DB}
\end{equation}
The diffused relativistic electrons and positrons generated by $\phi$ decay will scatter with the CMB and starlight photons. 
Note that a spatially independent magnetic field is needed when calculating Green’s function for the diffusion equation. 
In the integrand of Eq.~\eqref{ED}, we substitute the radial magnetic field strength $B(r)$ with the average value $B_{\rm avg}=\int B(r)dr/r$. In this work, $B_{\rm avg}$ is very close to the galactic magnetic field $B_{\rm G}$, and the difference can be negligible.

\subsection{Dark matter inverse Compton emission}
\label{sec:ICS}

Ref.~\cite{Orlando:2008uk} used the anisotropic Klein-Nishina scattering cross section when calculating the ICS from the Sun. 
Since the distance between Sirius A and the Earth is much further than one a.u., we follow Ref.~\cite{McDaniel:2017ppt} to assume that the ICS between the DM-induced electrons and the photons of Sirius A is isotropic. 
Furthermore, according to Ref.~\cite{Orlando:2008uk}, taking the angular distances $0.5^{\circ}$, 
the maximum difference of the ICS intensity from the Sun for two Klein-Nishina cross-sections is less than 15\%. 
For the Sirius system, such a small difference owing to anisotropic scattering can be neglected because two stars can be still treated as a point source from the observation on Earth.

The ICS power $\mathcal{P}_{\rm IC}$ is determined by photon number density $n_\gamma(\epsilon)$ and ICS scattering cross section $\sigma_{\gamma e}(\epsilon,E_\gamma,E_e)$ as  
\begin{equation}
\mathcal{P}_{\rm IC}(E_\gamma,E_e)=cE_\gamma\int_{\epsilon_{\rm min}}^{\epsilon_{\rm max}} d\epsilon~ n_\gamma (\epsilon)
\sigma_{\gamma e}(\epsilon,E_\gamma,E_e),
\label{ICpower}
\end{equation}
where $E_e$ is the energy of the relativistic electron and positron, 
but $E_\gamma$ is the energy of the upscattered photon.  
The energy of the target photon, minimum, and maximum energies are marked as
$\epsilon$, $\epsilon_{\rm min}$ and $\epsilon_{\rm max}$, respectively. 
The photon number density $n_\gamma(\epsilon)$ is given by the black body radiation formula  
\begin{equation}
n_\gamma(\epsilon)=\frac{8\pi\nu^2}{c^3}\frac{1}{e^{\epsilon/kT}-1}.
\label{blackbody}
\end{equation}
In this work, we consider the CMB and starlight photons from Sirius A as the target photons of the ICS.
In addition, we take the sum of these two parts as the total ICS. 
The cross-section $\sigma_{\gamma e}(\epsilon,E_\gamma,E_e)$ is given by the Klein-Nishina formula 
\begin{equation}
\sigma_{\gamma e}(\epsilon,E_\gamma,E_e)=\frac{3\sigma_{\rm T}}{4\epsilon\gamma_{e}^2}G(q,\Gamma),
\label{ICsigma}
\end{equation}
where $\sigma_{\rm T}$ is the Thomson cross-section, $\gamma_{e}=E/m_e$ is the Lorenz boost of the electron and $G(q,\Gamma)$ can be written as~\cite{Blumenthal:1970gc}
\begin{equation}
G(q,\Gamma)=\left[2q\ln{q}+(1+2q)(1-q)+\frac{(2q)^2(1-q)}{2(1+\Gamma q)}\right] 
\label{eq:G}
\end{equation}
with 
\begin{equation}
\Gamma=\frac{4\epsilon\gamma_{e}}{m_ec^2}=\frac{4\gamma_{e}^2\epsilon}{E_e}, \qquad q=\frac{E_\gamma}{\Gamma(E_e-E_\gamma)}.
\label{Gammaq}
\end{equation}
Note that $q$ has a range $1/(4\gamma_{e}^2)\leq q \leq 1$.

Finally, we can obtain the ICS local emission by integrating the diffused electron number density and ICS power as
\begin{equation}
j_{\rm IC}(E_\gamma,r)=2\int_{m_e}^{E_{\rm max}} dE_e \frac{dn_e}{dE_e}(E_e,r) \mathcal{P}_{\rm IC}(E_e,E_\gamma),
\label{Gammaq}
\end{equation}
and the photon flux received by the detector at Earth is
\begin{equation}
S_{\rm IC}(E_\gamma)=\frac{1}{D_{\rm A}^2}\int j_{\rm IC}(E_\gamma,r) r^2 dr.
\label{Gammaq}
\end{equation} 
In the consistency of the notation used in this paper, we rewrite the corresponding photon spectrum as
\begin{equation}
E_{\gamma}^2\frac{d\Phi_\gamma}{dE_\gamma}=\nu S_{\rm IC}(E_{\gamma}). 
\end{equation}

\subsection{Dark matter synchrotron emission}
\label{sec:syn}
The production of radio radiation occurs as a result of the synchrotron process 
when relativistic electrons and positrons traverse through a magnetic field. Compared to $\mathcal{P}_{\rm IC}$ obtained in the previous section, 
the synchrotron power $\mathcal{P}_{\rm syn}$ for a frequency $\nu$ averaged over all direction is written as~\cite{McDaniel:2017ppt,Storm:2016bfw}
\begin{equation}
\mathcal{P}_{\rm syn}(\nu,E_e,r)=\int_{0}^{\pi} d\theta \frac{\sin{\theta}}{2} 2\pi \sqrt{3} r_0 m_e c \nu_0 \sin{\theta} F(\frac{x}{\sin{\theta}}),
\label{Psyn}
\end{equation}
where $\nu_0=eB/(2\pi m_e c)$ is the non-relativistic gyrofrequency, $r_0=e^2/(m_e c^2)$ is the classical electron radius, 
and $\theta$ is the pitch angle. 
The $x$ and $F$ are given by 
\begin{equation*}
x=\frac{2\nu(1+z)m_e^2}{3\nu_0 E^2},~~{\rm and}~~
F(s)=s\int_{s}^{\infty} d\zeta K_{5/3}(\zeta) 1.25 s^{1/3} e^{-s} (648+s^2)^{1/12},
\end{equation*}
where $K_{5/3}$ is the Bessel function of order $5/3$. 
The synchrotron emissivity at the frequency $\nu$ can be obtained by 
integrating the diffused electron number density and the synchrotron power as
\begin{equation}
j_{\rm syn}(\nu,r)=2 \int_{m_e}^{E_{\rm max}} dE_e \frac{dn_e}{dE_e}(E_e,r) \mathcal{P}_{\rm syn}(\nu,E_e,r).
\end{equation}
Given the fact that the emission region is a small region compared to its distance to Earth, the detected frequency-dependent photon flux at Earth is then given by~\cite{Colafrancesco:2006he,Colafrancesco:2005ji,Beck:2015rna} 
\begin{equation}
S_{\rm syn}(\nu)=\frac{1}{D_{\rm A}^2} \int j_{\rm syn}(\nu,r) r^2 dr. 
\end{equation}
The corresponding photon spectrum then can be written as 
\begin{equation}
E_{\gamma}^2\frac{d\Phi_\gamma}{dE_\gamma}=\nu S_{\rm syn}(\nu). 
\end{equation}

\section{Total DM-induced photon spectrum}
\label{sec:tot_spectra}

In this section, we demonstrate the total DM-induced photon fluxes 
(including the contributions from the prompt photon emission, the ICS with CMB and Starlight, 
and the synchrotron emission) based on the four benchmarks given in the left panel of Fig.~\ref{fig:ga_tau}. 
For the ICS and prompt photon emission, we neglect the contribution from Sirius B. While for the synchrotron emission, we consider the contribution from Sirius A and Sirius B. As a comparison, we use the Fermi-LAT data in the gamma-ray frequency  
but the future SKA sensitivities in the radio frequency.   

First, we compare the expected photon spectra of the ICS and prompt photon emission in the gamma-ray energy range with the Fermi-LAT observation in the direction of Sirius. 
We use more than 13 years of Fermi-LAT P8R3 data from October 27, 2008 to February 01, 2022. 
We select photons from 300 MeV to 1 TeV in the {\tt SOURCE} event class with the {\tt FRONT+BACK} conversion type.
To reduce the contamination from the Earth’s limb and the Sun, we exclude photons with zenith angles larger than $90\rdeg$ 
and extract good time intervals with the quality-filter cut 
{\tt (DATA\_QUAL==1 \&\& LAT\_CONFIG==1) \&\& (angsep(ra, dec, RA\_SUN, DEC\_SUN)>15)}.

In our analysis, the standard binned likelihood method is used 
in a $14\rdeg \times 14\rdeg$ region of interest (ROI) centered on the location of Sirius (ra, dec) = ($101.3\rdeg$, $-16.7\rdeg$). 
The {\tt Fermitools} package~\footnote{\url{https://github.com/fermi-lat/Fermitools-conda/}} and 
the {\tt P8R3\_SOURCE\_V3} instrument response functions are utilized in our work.
We generate the initial model for the ROI region with the {\tt make4FGLxml.py} script. 
This model includes the galactic diffuse emission template ({\tt gll\_iem\_v07.fits}), 
the isotropic diffuse spectral model for the {\tt SOURCE} data ({\tt iso\_P8R3\_SOURCE\_V3\_v1.txt}) and 
all the Fourth Fermi-LAT source catalog~\cite{1902.10045} ({\tt gll\_psc\_v30.fit}) source within $15\rdeg$ of Sirius. 
We model the gamma-ray emission from Sirius as a point source and set its spectral shape 
to the {\tt PowerLaw} model ($\propto E^\alpha$ with the spectral index $\alpha$). 
In the fitting process, we vary normalizations and spectral indexes of all sources within $7\rdeg$ 
from Sirius and normalizations of the Galactic diffuse and isotropic diffuse components.

Fitting this model with the Fermi-LAT data,
we obtain the test statistic (TS) value of Sirius is 1.1. 
Hence, no statistically significant gamma-ray emission in the direction of Sirius has been detected in our analysis. 
Then, we use {\tt UpperLimits} tool in the {\tt Fermitools} package to calculate 
the 95\% upper limit of the flux from Sirius with the spectral index of $\alpha=-2$.
The 95\% upper limit we obtained is $2.01\times 10^{-10} \, {\rm ph~cm^{-2}s^{-1}}$ within the energy range from 300 MeV to 1 TeV, which is used to constrain DM parameter space via the ICS and prompt radiation processes in the following section.

Next, in the radio frequency band, we compare our results to the future SKA of frequencies ranging from 50 MHz to 50 GHz. 
The SKA is a next-generation radio telescope array, consisting of thousands of individual radio antennas spreading across two countries, South Africa and Australia~\cite{Braun:2019gdo}. 
The first phase of SKA (SKA1) with a frequency range from 50 MHz to about 15 GHz, including SKA1-Mid and SKA1-Low, has been under construction since 2021.
Because of its high sensitivity and energy resolution,      
SKA can soon set sensitivities of radio emission 
around $\mathcal{O}(\mu {\rm Jy})$ ($\sim 10^{-29}$ erg$\cdot$cm$^2\cdot s^{-1}\cdot$Hz$^{-1}$)~\cite{Dewdney:2013ard}. 
The SKA sensitivities have been used to investigate for signals from the DM-induced synchrotron~\cite{Colafrancesco:2014coa,Beck:2021xsv,Kar:2018rlm} and ICS emission~\cite{Dutta:2020lqc}.
In this work, we adopt the same approach in our previous work~\cite{Chen:2021rea}. 
The minimum detectable flux of SKA depends on the ratio of the effective collecting area $A_e$ and the total system noise temperature $T_{\mathrm{sys}}$~\cite{Braun:2019gdo}, which can be given as
\begin{equation}
S_{\mathrm{min}}=\frac{2k_{b}S_{D}T_{\mathrm{sys}}}{\eta_{s}A_e(\eta_{\mathrm{pol}}\mathcal{T}\Delta\nu)^{1/2}},
\label{SKA_sensitivities}
\end{equation}
where $k_{b}$ is the Boltzmann constant, $\eta_{s}$ is the system efficiency, $\eta_{\mathrm{pol}}$ is the number of polarisation states, $\mathcal{T}$ represents the total integration time, $S_{D}$ is a degradation factor for the noise in a continuum image and is denoted as $S_{D}=2$. We set the channel bandwidth $\Delta\nu$ as 0.3 times of the frequency. 
To study on the DM-induced synchrotron emission, here we adopt the first and second phases of SKA (SKA1 and SKA2) sensitivities with the exposure time of $\mathcal{T}=100$ hours.

\begin{figure*}
\centering 
\includegraphics[width=0.49\textwidth]{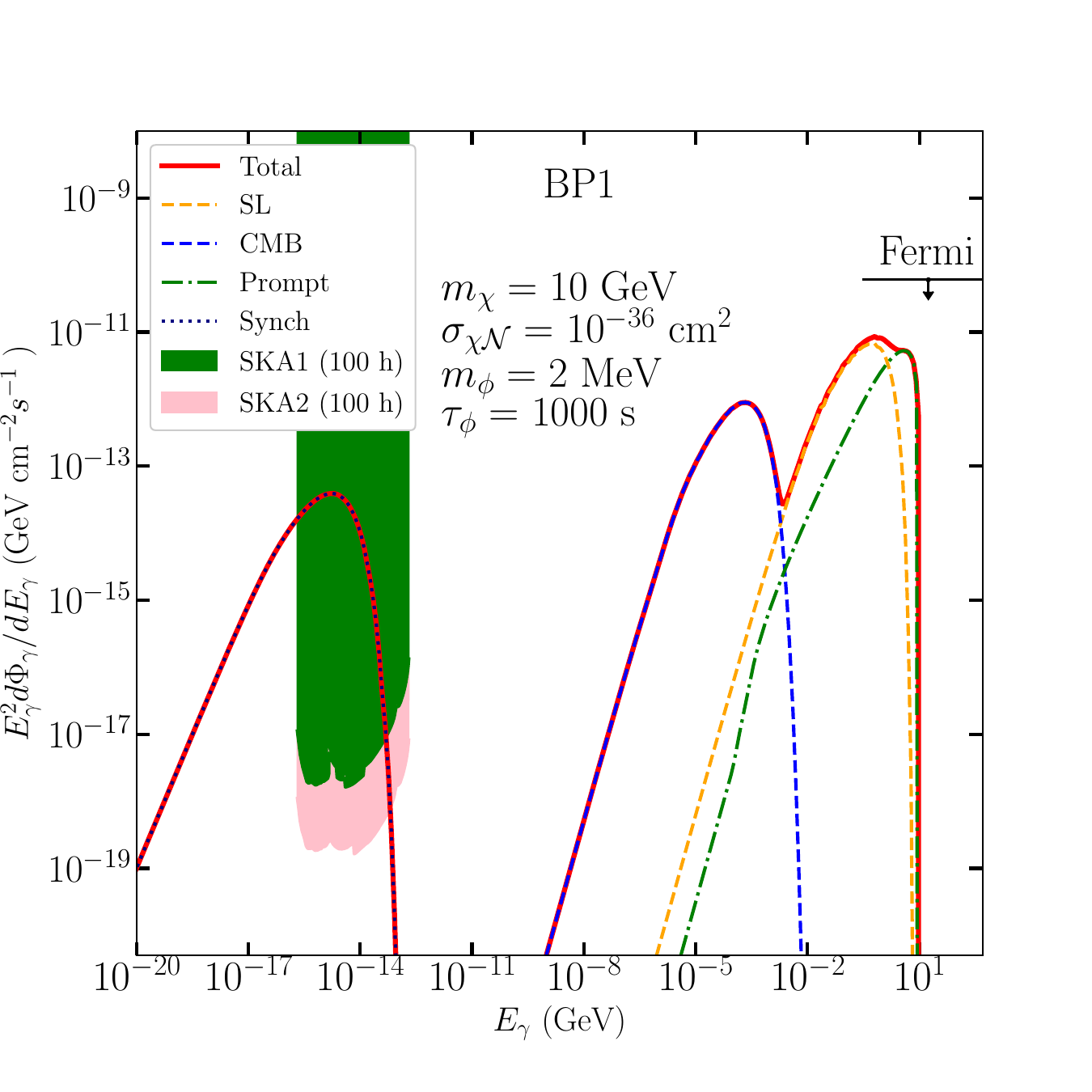}
\includegraphics[width=0.49\textwidth]{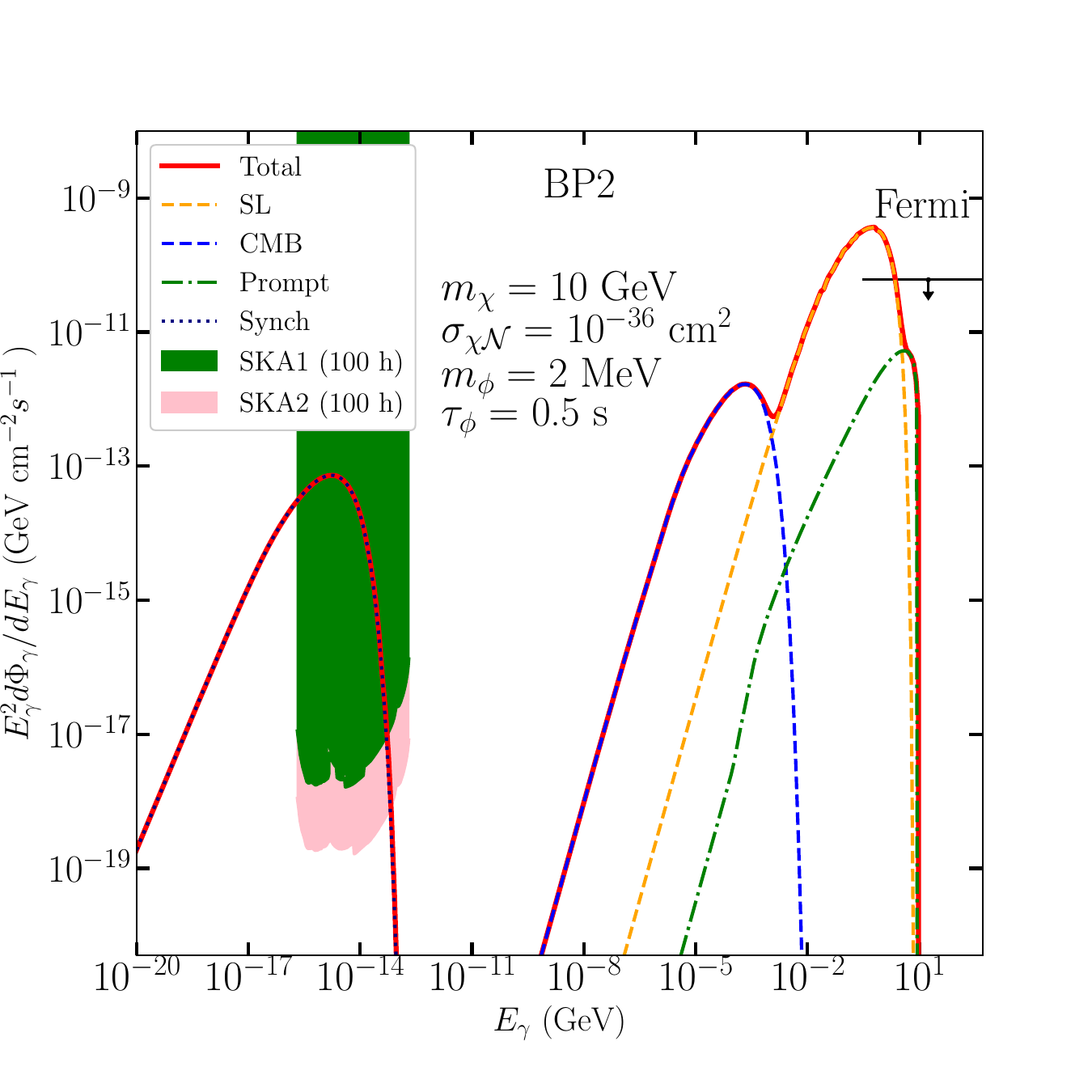}
\includegraphics[width=0.49\textwidth]{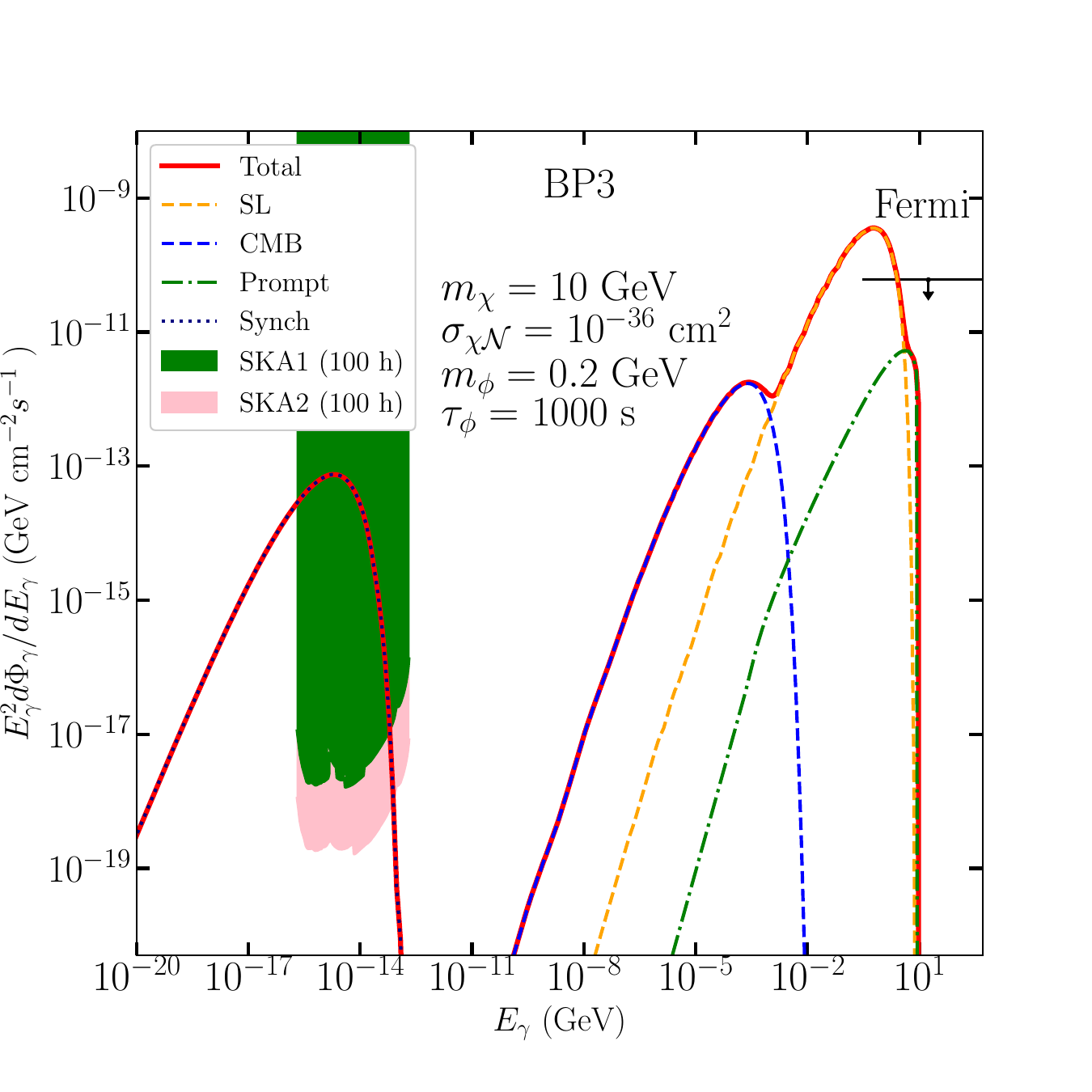}
\includegraphics[width=0.49\textwidth]{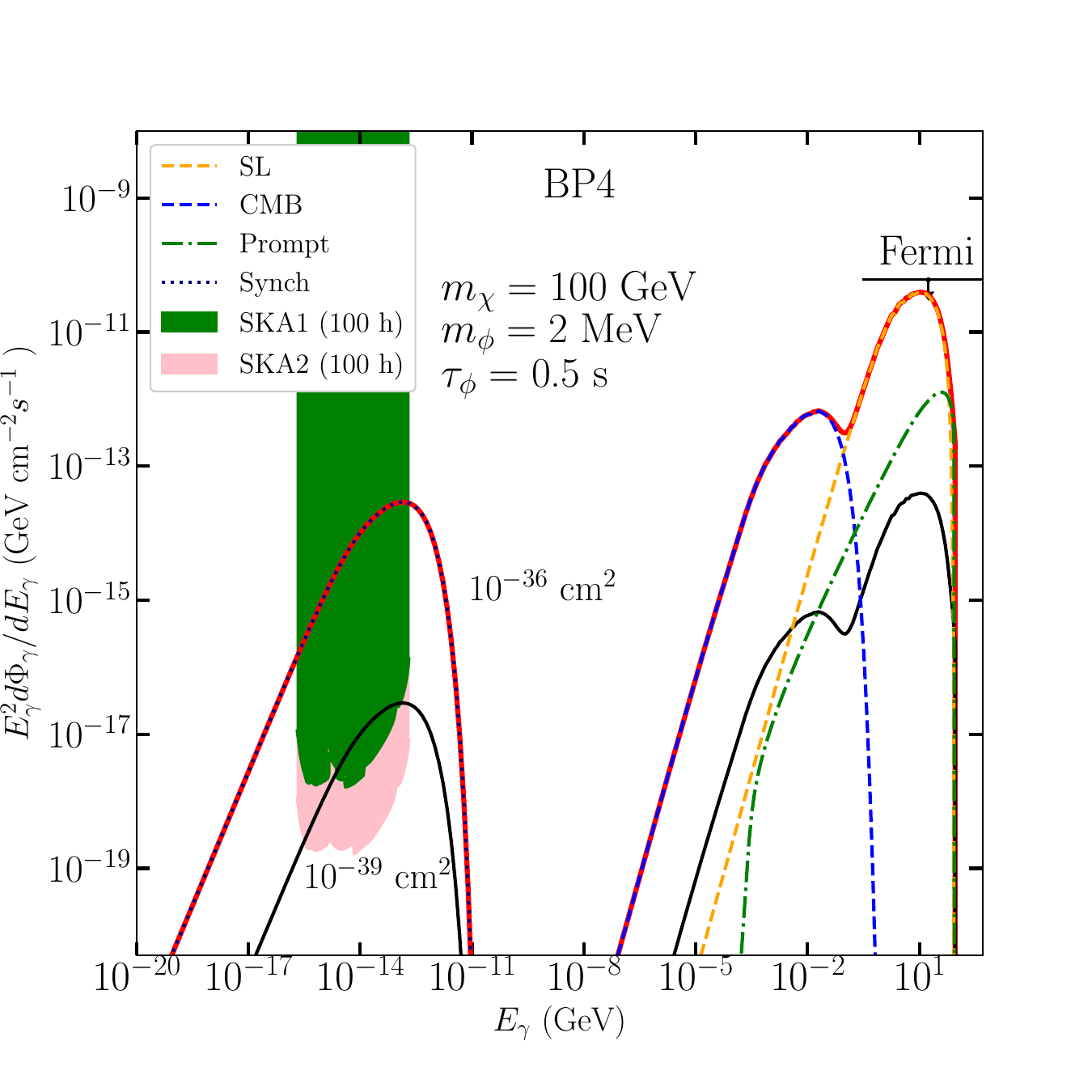}
\caption{The energy spectra as a function of the photon energy for benchmarks given 
in the left panel of Fig.~\ref{fig:ga_tau}.
The solid lines indicate the total DM-induced photon fluxes, including the contributions of 
prompt production (dash-dotted lines), ICS (dashed lines), and synchrotron radiation (dotted lines). 
In the lower right panel, two values of cross section 
(red solid line for $\sigxn=10^{-36}~\rm cm^2$ and black solid line for $\sigxn=10^{-39}~\rm cm^2$) 
are compared. 
The expected sensitivity of SKA1 and SKA2~\cite{Chen:2021rea} 
are shown as the green region and pink region, respectively.
} 
\label{Fig:Flux} 
\end{figure*}

Fig.~\ref{Fig:Flux} shows the energy spectra as a function of the photon energy for 
BP1 (upper-left panel), BP2 (upper-right panel), BP3 (lower-left panel), and 
BP4 (lower-right panel). Comparing BP1 and BP2, only $\tau_\phi$ is changed from $10^3$~s to $0.5$~s, 
but $\phi$ with a shorter lifetime could decay closer to the star.
Thus, the generated $e^\pm$ interacts with the higher intensity of starlight and 
produces a larger gamma-ray flux via ICS, seen the orange lines in the upper two panels.  
Even though $\tau_\phi$ of BP3 is also $10^3$, the ICS with starlight is comparable with BP2 and 
much larger than BP1, due to a similar $\gamma_\phi c\tau_\phi$ as illustrated in Fig.~\ref{fig:ga_tau}.

In the lower-right panel of Fig.~\ref{Fig:Flux}, we can see that the total photon fluxes of BP4 are also lower than BP2 even if their $\phi$ parameter configurations are identical. 
The only difference between BP2 and BP4 is the DM mass which causes a lower capture rate reflected in the overall spectra. In addition to $\sigxn=10^{-36}$~cm$^2$, 
we also present a case $\sigxn=10^{-39}$~cm$^2$ (black solid line) as a comparison. 
We can see that the photon fluxes cannot be linearly scaled with respect to $\sigxn$.


Here are some comments on the impact of the diffusion coefficient $D_0$ and the galactic magnetic field $B_{\rm G}$ on the total photon fluxes. 
We vary $D_0/({\rm cm}^{2}{\rm s}^{-1})$ between $10^{27}$ and $10^{29}$, 
but find that changes in the photon energy spectra are negligible. 
This can be understood that the diffusion scale corresponding to $D_0=10^{28}~\rm{cm^2~s^{-1}}$ 
is larger than the distance from Sirius to the Earth. 
On the other hand, we also numerically confirm that 
the total photon spectra can be slightly altered if 
we vary the galactic magnetic field within the range $1<B_{\rm G}/\mu {\rm G}<4$. 
Hence, we simply neglect the systematic uncertainties from $D_0$ and $B_{\rm G}$ in this work.

\begin{figure*}[!ht]
\centering 
\includegraphics[width=10cm]{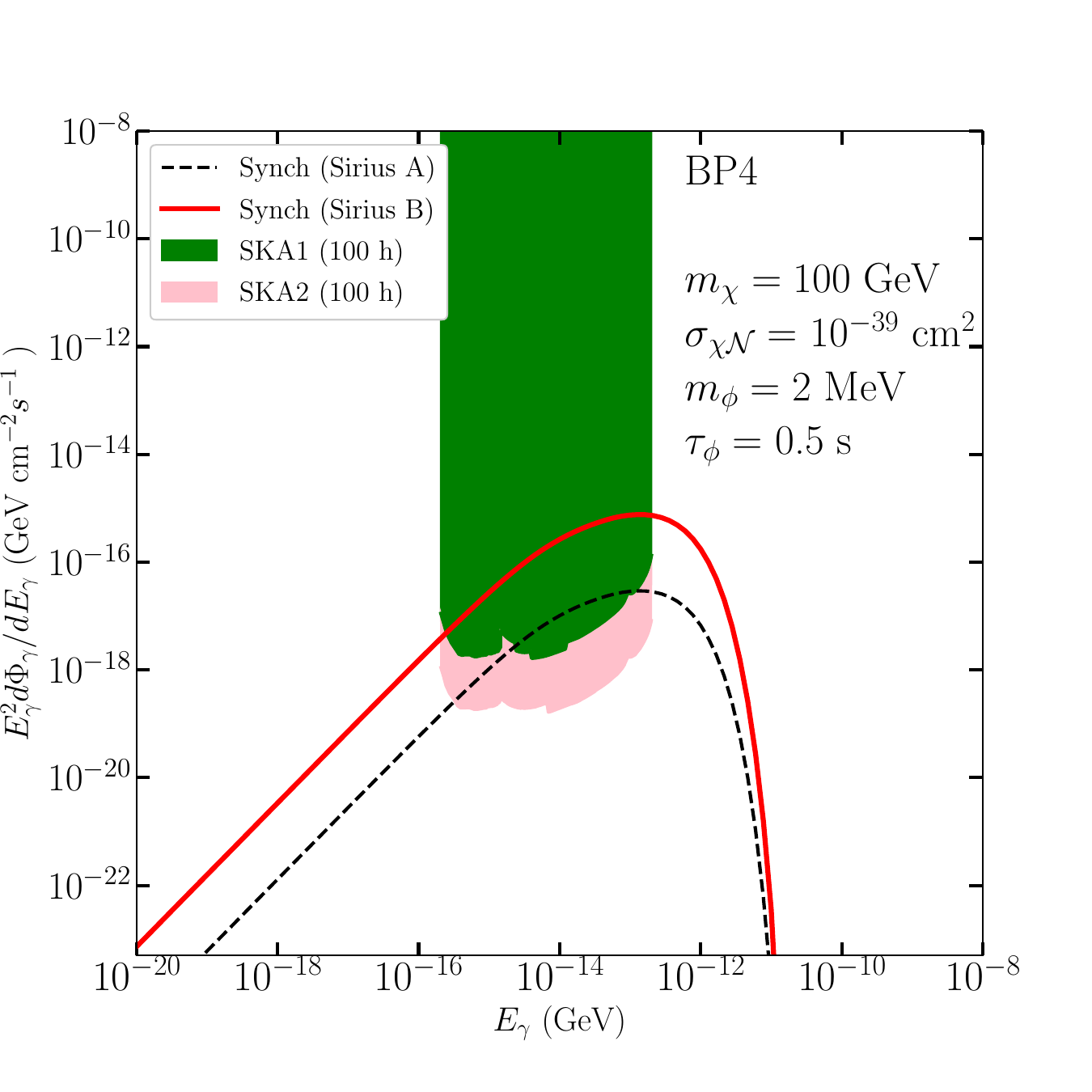}
\caption{The energy spectra of synchrotron radiation for Sirius A (dashed black line) and Sirius B (red line).} 
\label{Fig:Flux_Sigma2} 
\end{figure*}

Fig.~\ref{Fig:Flux_Sigma2} shows that the photon energy spectrum for Sirius A (red line) and 
Sirius B (dashed black line) in radio frequency. 
Note that the black dashed line in Fig.~\ref{Fig:Flux_Sigma2} is actually the same as the black solid line in the lower-right panel of Fig.~\ref{Fig:Flux}. We see that the DM-induced photon flux from Sirius B is about two order magnitudes larger than 
the flux from Sirius A. 
This is due to two combinatorial facts. 
First, the DM capture rate of Sirius B 
is higher than that of Sirius A as shown in Fig.~\ref{Fig:Capture_Rate}. 
Second, the surface magnetic field of Sirius B is overwhelmingly stronger than that of Sirius A. Note that we take $B_0^{\rm SA}=0.2$~G and $B_0^{\rm SB}=1$~MG in this work, 
and synchrotron radiation is more sensitive to magnetic fields.

\section{Result} 
\label{sec:result}

In this section, we project the current Fermi limits and future SKA sensitivity on 
($m_\chi$, $\sigsip$) and ($m_\chi$, $\sigsdp$) planes. 
In principle, both spin-dependent and spin-independent components can contribute to the DM capture rate $C_{\rm cap}$.  
However, the scattering between DM and the primary elements of Sirius A (Hydrogen) differs from 
its scattering with the primary elements of Sirius B (Carbon). 
Both $\sigsip$ and $\sigsdp$ contribute to $C_{\rm cap}$ of Sirius A. 
In contrast, only $\sigsip$ is important to $C_{\rm cap}$ of Sirius B because 
the zero nuclear spin of the Carbon nucleus leads to the vanished spin-dependent component of DM-Carbon scattering,
as outlined in Eq.~\eqref{eq:si_sd}.
Therefore, we present $\sigsip$ and $\sigsdp$ separately as the DM DD experiments do. 
When deriving the limits on $\sigsip$, we consider the DM-induced photons generated from both Sirius A and Sirius B. 
While for the limits on $\sigsdp$, we only consider the DM-induced photons generated from Sirius A.

\begin{figure*}[!ht]
\centering 
\includegraphics[width=10cm]{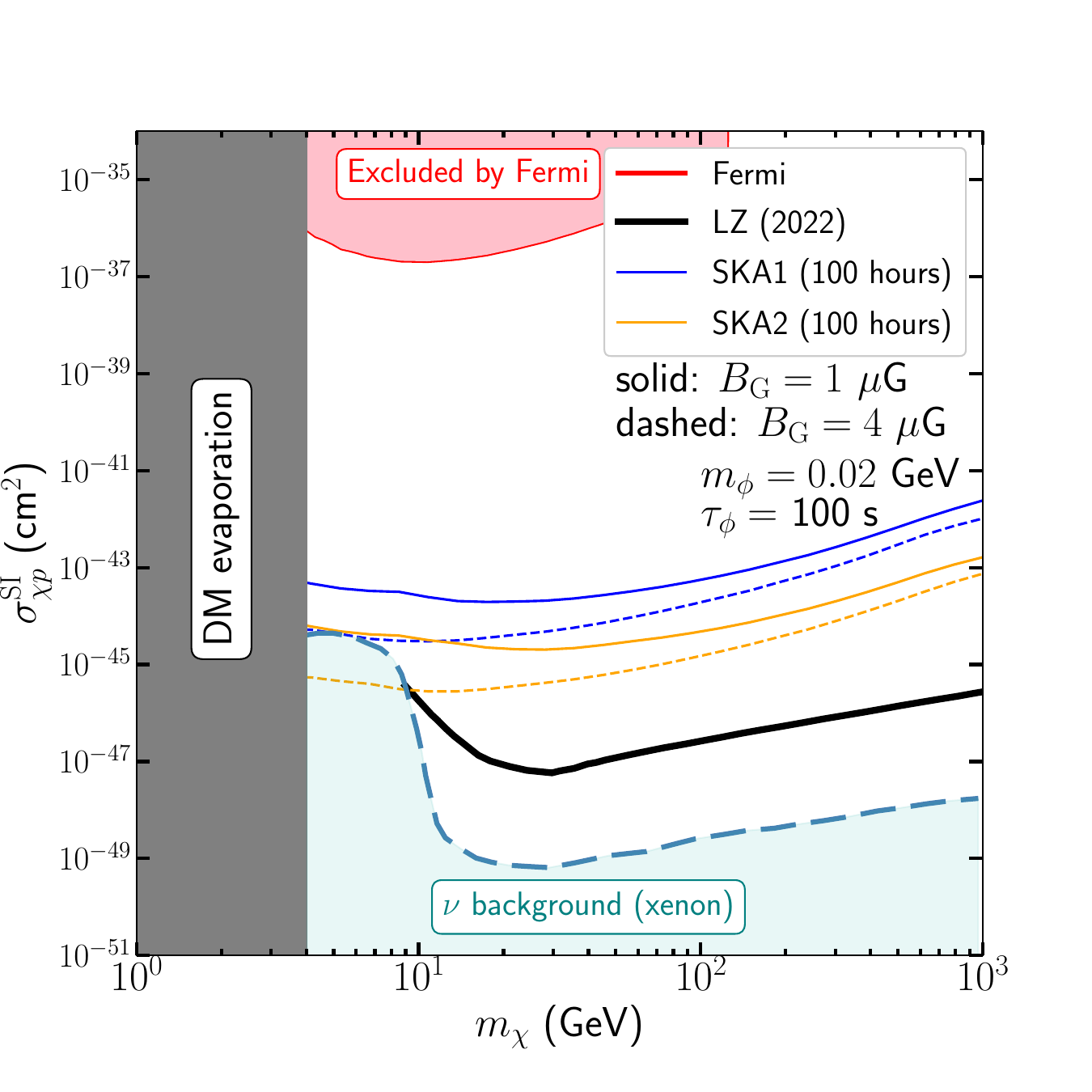}
\caption{Spin-independent DM-proton scattering cross section limits as a function of DM mass $m_\chi$,  
assuming $m_\phi=0.02\gev$ and the lifetime of $\phi$ is $\tau_\phi = 100$ s. 
The pink region is the $95\%$ exclusion of the Fermi-LAT when taking the ICS and prompt contribution into account. 
The gray region corresponds to the DM evaporation dominant region 
where DM is heated to escape from the gravitational potential. The cyan region is the neutrino floor for xenon targets.~\cite{OHare:2021utq}
The SKA1 (blue lines) and SKA2 (orange lines) sensitivities are compared with the LUX-ZEPLIN (LZ) limits~\cite{LZ:2022ufs} (black lines). 
The solid and dashed lines are for the galactic magnetic field strength of 
$B_{\rm G} = 1~\mu {\rm G}$ and $4~\mu {\rm G}$, respectively.}
\label{Fig:Constrain_SI} 
\end{figure*}

In Fig.~\ref{Fig:Constrain_SI}, we derive the 95\% Fermi-LAT exclusion of $\sigsip$ 
as presented by pink regions, via $\chi\chi \rightarrow \phi\phi \rightarrow 4e$. For parameters of the long-lived particles, we fix the mass  $m_\phi=0.02\gev$ and a long lifetime $\tau_\phi=100$~s. 
The decay length $\gamma_\phi c \tau_\phi<3\times10^{15}~\rm{cm}$ leads the strongest limit as shown in Fig.~\ref{fig:ga_tau}. We have demonstrated that the limit keeps unchanging relatively within the range of $\mathcal{O}(1)$ s to $\mathcal{O}(100)$ s for $\tau_\phi$ through numerical calculations.
The gray region indicates where DM can be evaporated after a DM-nucleus scattering. The cyan region indicates the neutrino floor for xenon targets~\cite{OHare:2021utq}.
For a given DM mass within the range $4\gev<m_\chi<100\gev$, 
we vary $\sigsip$ to obtain the upper limit when the expected photon fluxes for  
the sum of ICS and prompt components reach the Fermi-LAT upper limit. 
We note that the DM mass regions $m_\chi>100\gev$ correspond to where 
the Fermi-LAT sensitivity is no longer stronger than the geometric limit.

As for the synchrotron radiation, we use the expected 100 hours of sensitivity of 
the upcoming SKA 1 (blue lines) and SKA 2 (orange lines). 
We take a conservative approach to the expected DM detection sensitivity, by requiring the photon flux 
of the synchrotron radiation not to exceed the SKA sensitivity over the whole range of observed frequency.  
We also compare two representative values of the galactic magnetic field, 
$B_{\rm G} = 1~\mu {\rm G}$ (solid lines) and $4~\mu {\rm G}$ (dashed lines). We found that the expected sensitivity keeps relatively stable within the range of $\mathcal{O}(1)$ s to $\mathcal{O}(100)$ s for $\tau_\phi$.

Compared with the limits from the LZ experiment~\cite{LZ:2022ufs} (black lines), 
the latest Fermi-LAT gamma-ray data cannot probe $\sigsip$ as powerfully as the DM DD experiment. However, the prospective sensitivity of $\sigsip$ given by SKA can be significantly improved.  
Especially in the $\mathcal{O}(\gev)$ DM regime, due to the shortage of DM recoil energy below the LZ detector threshold, 
the sensitivity of SKA2 can be more stringent than the current LZ limits.

\begin{figure*}[!ht]
\centering 
\includegraphics[width=10cm]{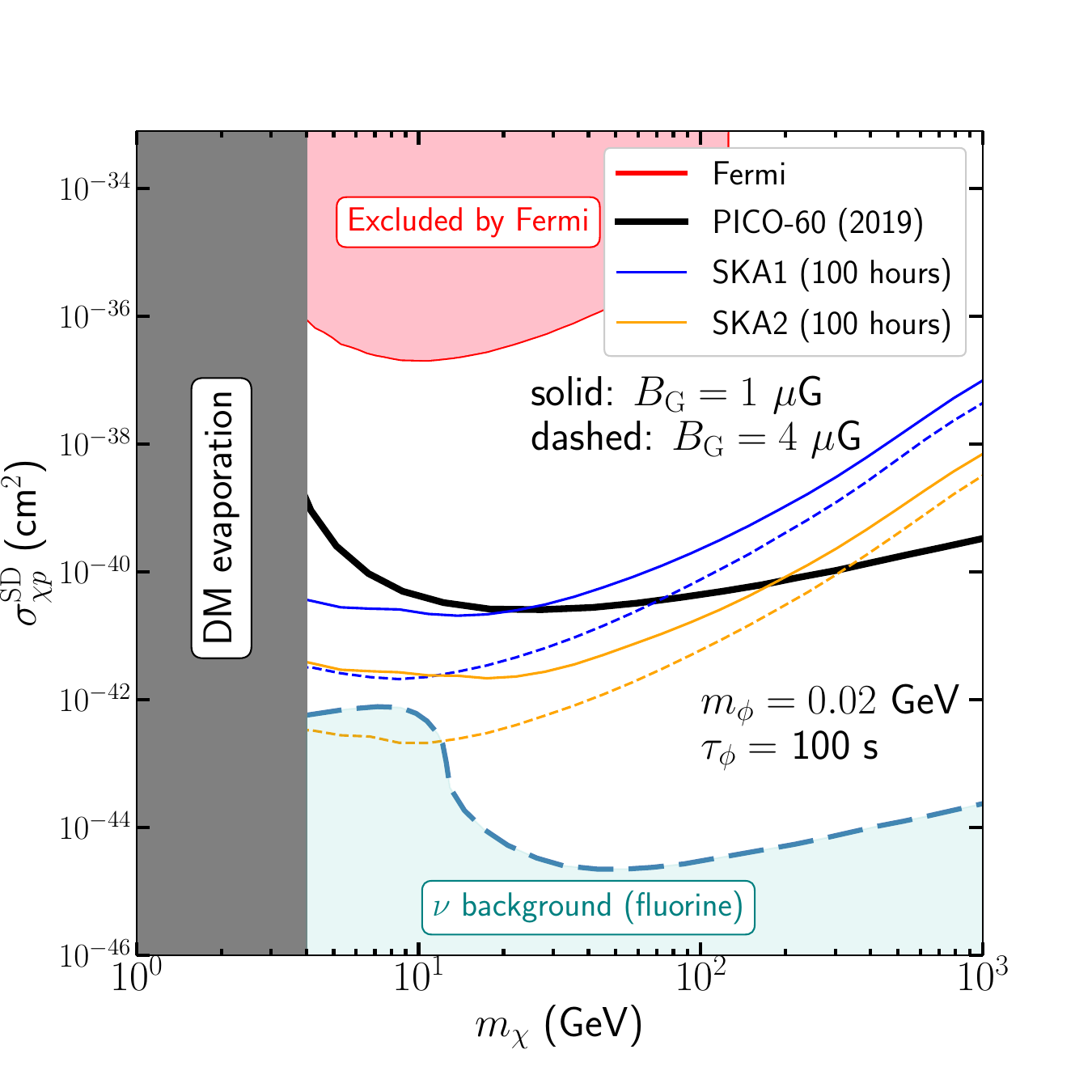}
\caption{The spin-dependent DM-proton scattering cross section limits as a function of DM mass $m_\chi$. 
The color coding in the shaded regions is the same as in Fig.~\ref{Fig:Constrain_SI}, 
except for the PICO-60 DD limits~\cite{PICO:2019vsc} (black lines) and the neutrino floor for fluorine targets (cyan region)~\cite{OHare:2021utq}.
}
\label{Fig:Constrain} 
\end{figure*}

In Fig.~\ref{Fig:Constrain}, we show the $\sigsdp$ exclusion by using the same method applied in Fig.~\ref{Fig:Constrain_SI}. 
The most stringent limits of $\sigsdp$ (black line) are taken from the PICO-60~\cite{PICO:2019vsc} and the neutrino floor is for fluorine targets~\cite{OHare:2021utq}. 
Even though there is still a gap between two current limits (Fermi-LAT and PICO-60), 
it is much smaller than the gap between Fermi-LAT and LZ in Fig.~\ref{Fig:Constrain_SI}.  
In Fig.~\ref{Fig:Constrain}, we can see that 
the future SKA1 and SKA2 sensitivities can probe the $m_\chi<100\gev$ region and exceed PICO-60 limits by 2 orders of magnitude (SKA1) and 3 orders of magnitude (SKA2). In sum, the upcoming SKA would provide another window to probe DM annihilation to a pair of long-lived particles.

\section{Summary and conclusion} 
\label{sec:summary}

In this paper, we study the photon signals generated by long-lived particles with a focus on their decay length longer than the distances from Earth to the Sun and Jupiter.
Considering DM from the galactic halo captured by the gravitational potential of stars, 
such long-lived particles can be produced by the star-captured DM annihilation and 
$e^\pm$ yielded by their subsequent decay can generate photons via (i) the final state radiation (prompt component), 
(ii) the ICS with the starlight and CMB, and 
(iii) the synchrotron emission from the interaction with the galactic and star magnetic field. 
In our quest for an ideal target to identify photons generated by DM, we have examined the Sirius system, which is comprised of two stars---Sirius A and Sirius B. 
This marks the first systematic investigation of DM annihilation into long-lived particles in the Sirius system. 
Sirius A is primarily composed of Hydrogen, which makes its DM capture rate sensitive to both $\sigsdp$ and $\sigsip$. 
On the other hand, Sirius B, which is primarily made up of Carbon, has a DM capture rate that is only sensitive to $\sigsip$.
Due to the complex photon emission, we analyze photon fluxes in the multifrequency:
one is gamma rays from the prompt component and ICS based on the latest Fermi-LAT data, 
and the other is synchrotron emission in the radio band based on the future SKA prospect.

Our calculations have determined the overall energy spectra of DM-induced photons from the Sirius system, 
including prompt, Inverse Compton Scattering, and synchrotron emissions. 
Using comprehensive formulas of DM-induced photon fluxes, we have established the upper limits from the Fermi-LAT observations as well as the expected sensitivity with the future SKA telescope. 
Our results demonstrate that although the upper limits on $\sigsip$ and $\sigsdp$ by Fermi-LAT may not be as stringent as those from DM DD experiments, the future SKA sensitivity on $\sigsdp$ could surpass the limits from PICO-60. 
Our findings are dependent on the decay length of the mediator $\gamma_\phi c\tau_\phi$. 
For instance, for $m_\phi=20\mev$ and $\tau_\phi=100$~s, the SKA could detect $\sigsdp$ with 2-3 orders of magnitude greater sensitivity compared to PICO-60 in the mass range $m_\chi\lesssim 100\gev$. 
Our analysis also found that the impact of the galactic magnetic field $B_{\rm G}$ on the synchrotron emission could be more pronounced than that of the diffusion coefficient $D_0$.

\begin{equation}
\frac{d\Phi_{\gamma}}{dE_\gamma}=\frac{1}{4 \pi D^2} \int_{V_{\rm {eff}}} n_a^{\rm {orb}} \Gamma_{a \gamma \gamma} d \overline{r}. 
\end{equation}

\begin{equation}
1<\overline{r}<10
\end{equation}

\begin{acknowledgments}
This work is supported by the National Key Research and Development Program of China (No. 2022YFF0503304), the National Natural Science Foundation of China (No. 11921003, No. 12003069, No. U1738210) and the Entrepreneurship and Innovation Program of Jiangsu Province.
\end{acknowledgments}

\end{document}